\documentclass[twocolumn,aps,preprintnumbers,amsmath,amssymb,superscriptaddress,nofootinbib]{revtex4}

\usepackage{graphicx,subfigure,multirow}
\usepackage{longtable}
\usepackage{dcolumn}
\usepackage{bm}
\usepackage[all]{xy}
\usepackage{color}
 \usepackage{graphicx}
\usepackage[usenames,dvipsnames]{xcolor}
\usepackage[unicode=true,pdfusetitle,
 bookmarks=true,bookmarksnumbered=false,bookmarksopen=false,citecolor=Turquoise,
 breaklinks=false,pdfborder={0 0 1},backref=false,colorlinks=true,pdfpagemode=FullScreen]
 {hyperref}

\usepackage{lscape}
\usepackage{tikz}
\usepackage{pgf}
\usepackage{booktabs}
\usepackage{color}

\usepackage{lipsum}

\makeatletter

\newcommand{\fmslash}[2][0mu]{%
  \mathchoice
    {\fmsl@sh\displaystyle{#1}{#2}}%
    {\fmsl@sh\textstyle{#1}{#2}}%
    {\fmsl@sh\scriptstyle{#1}{#2}}%
    {\fmsl@sh\scriptscriptstyle{#1}{#2}}}
\newcommand{\fmsl@sh}[3]{%
  \m@th\ooalign{$\hfil#1\mkern#2/\hfil$\crcr$#1#3$}}
\makeatother

\newcommand{\beq}{\begin{equation}}
\newcommand{\eeq}{\end{equation}}
\newcommand{\bea}{\begin{eqnarray}}
\newcommand{\eea}{\end{eqnarray}}
\mathchardef\minus="002D

\newcommand{\CLASS}{\texttt{CLASS}}

\begin{document}
\title{Warm Surprises from Cold Duets: \\
$N$-Body Simulations with Two-Component Dark Matter}

\author{Jeong Han Kim}
\email{jeonghan.kim@cbu.ac.kr}
\affiliation{Department of Physics, Chungbuk National University, Cheongju, Chungbuk 28644, Republic of Korea}

\author{Kyoungchul Kong}
\email{kckong@ku.edu}
\affiliation{Department of Physics and Astronomy, University of Kansas, Lawrence, KS 66045, USA}

\author{Se Hwan Lim}
\email{sehwan.lim@chungbuk.ac.kr}
\affiliation{Department of Physics, Chungbuk National University, Cheongju, Chungbuk 28644, Republic of Korea}

\author{Jong-Chul Park}
\email{jcpark@cnu.ac.kr}
\affiliation{Department of Physics and IQS, Chungnam National University, Daejeon 34134, Republic of Korea}

\begin{abstract}
We explore extensive $N$-body simulations with two-component cold dark matter candidates. 
We delve into the temperature evolution, power spectrum, density perturbation, and maximum circular velocity functions. 
We find that the substantial mass difference between the two candidates and the annihilation of the heavier components to the lighter ones effectively endow the latter with warm dark matter-like behavior, taking advantage of all distinct features that warm dark matter candidates offer, without observational bounds on the warm dark matter mass.
Moreover, we demonstrate that the two-component dark matter model aligns well with observational data, providing valuable insights into where and how to search for the elusive dark matter candidates in terrestrial experiments. 
\end{abstract}


\maketitle

\noindent {\bf Introduction.} 
Nature of dark matter (DM) remains mysterious even after decades of a multitude of experiments in particle physics and great advances in observational astrophysics and cosmology \cite{Cooley:2022ufh}. 
The majority of cosmological simulations focus on a single DM candidate for its simplicity and approximate consistency with much of observational data, while various theoretical models of dark matter, including multi-component dark matter~\cite{DEramo:2010keq, Belanger:2011ww}, have been proposed to address some of unresolved queries and to better understand data in current and upcoming experiments \cite{Perivolaropoulos:2021jda}. 
Most existing studies on two-component dark matter consider $N$-body simulation with non-interacting two cold dark matter (CDM) components \cite{Huang:2022ffc} or with non-interacting CDM and warm dark matter (WDM) \cite{Maccio:2012rjx, Parimbelli:2021mtp,Boyarsky_2009,Macci__2012,Anderhalden_2012}. 
A non-trivial connection between two-component dark matters and its cosmology has been discussed in literature, but the model requires 
extremely degenerate mass spectrum $\Delta m/m \sim 10^{-8}$ 
\cite{Medvedev:2013vsa,Todoroki_2018,Todoroki:2017pdh,Todoroki:2017kge,Todoroki:2020num}.

\begin{figure}[t!]
    \centering
 \includegraphics[width=1.0\linewidth,clip]{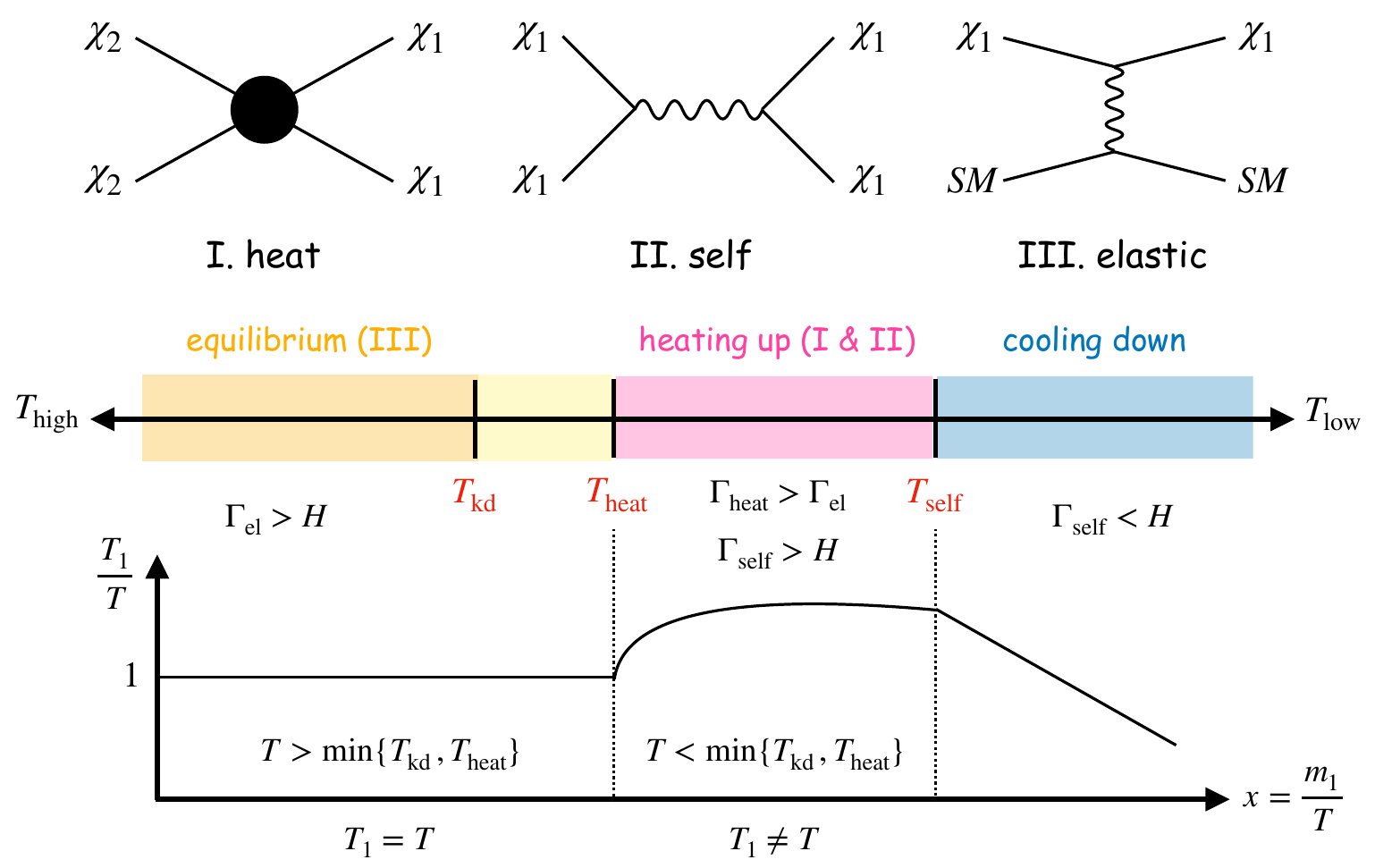}
 \vspace*{-0.3cm}
    \caption{Brief thermal history of the universe with two-component CDMs. 
    Different stages are defined by three important temperatures, $T_{\rm kd}$, $T_{\rm heat}$, and $T_{\rm self}$, which are determined by the balance between the various interaction rates and the Hubble parameter $H$:   
$\Gamma_{\rm el} \leftrightarrow H$, $\Gamma_{\rm el} \leftrightarrow \Gamma_{\rm heat}$, and $\Gamma_{\rm self} \leftrightarrow H$, respectively.
$\Gamma_{\rm heat}$, $\Gamma_{\rm self}$, and $\Gamma_{\rm el}$ are the interaction rates due to heating process (I), self-interaction (II), and elastic scattering (III), respectively. 
\label{fig:diagram}}
\end{figure}

In this paper, we go beyond one single CDM paradigm and perform comprehensive cosmological simulations for two-component CDM candidates with the sizable mass splitting and self-interaction (diagram II in Figure \ref{fig:diagram})  
for the light component. 
After both components decouple from the thermal bath, in addition to the heavy CDM, we expect two types of light CDMs to arise: relic (non-relativistic) component and boosted DM from the annihilation of the heavy CDM. 
The light-boosted components share their energy with the slow-relic component, increasing the temperature via sizable non-gravitational self-interaction, and therefore making the light relic behave like WDM, 
as illustrated in Figure \ref{fig:diagram}.
Such a scenario naturally connects between models with two CDMs and models with CDM + WDM. Therefore, this model innately takes advantage of all distinct features that WDM offers, such as power spectrum, dark matter density profile, small-scale structures, etc.
On the other hand, it suffers less from observational bounds on the WDM mass. 
For example, WDM mass bound from Lyman-$\alpha$ observation \cite{Garzilli:2019qki, Villasenor:2022aiy} is significantly weakened or negligible in this model. 
WDM becomes hotter as it gets lighter.
However, our model controls such behavior with the relic fraction of the light component to the total DM abundance and mass splitting.   

Many direct detection schemes have been proposed targeting sub-MeV DM \cite{Hochberg:2015pha,Schutz:2016tid,Hochberg:2017wce,Knapen:2017ekk,Hochberg:2019cyy,Kim:2020bwm}, while currently operating experiments are not sensitive to the typical WDM mass~\cite{SENSEI:2020dpa,Hochberg:2021yud}.
However, the boosted property of the light DM component offers great opportunity in the particle physics experiments \cite{Agashe:2014yua, Berger:2014sqa, Kong:2014mia, Cherry:2015oca, Necib:2016aez, Alhazmi:2016qcs, Kim:2016zjx, Super-Kamiokande:2017dch, Giudice:2017zke, Chatterjee:2018mej, Kim:2018veo, Bringmann:2018cvk, Ema:2018bih, COSINE-100:2018ged, Kim:2019had, Heurtier:2019rkz, Berger:2019ttc, Kim:2020ipj, DeRoeck:2020ntj, Cao:2020bwd, Jho:2020sku, Alhazmi:2020fju, Jho:2021rmn, Wang:2021nbf, Wang:2021jic, PandaX-II:2021kai, CDEX:2022fig, Bell:2021xff, COSINE-100:2023tcq} and cosmological thermal dynamics \cite{Kamada:2021muh}. 
Yet, little studies on cosmological implications of such boosted DM models have been discussed. 
In this paper, we will carefully examine the temperature evolution, density perturbation, and $N$-body simulation of the two-component DM model. 
We derive the model parameters that are consistent with the maximum circular velocity function (MCVF), which can be studied in current and upcoming particle physics experiments.


\medskip
\noindent {\bf Model.} 
We consider a two-component DM model with a dark sector consisting of two Dirac fermions $\chi_1$ and $\chi_2$ with mass hierarchy $m_2 > m_1$, whose stability is protected by dark ${\rm U}(1)'\otimes{\rm U}(1)''$ gauge symmetries~\cite{Belanger:2011ww}. 
We assume that both $\chi_1$ and $\chi_2$ are charged under ${\rm U}(1)''$, while only $\chi_1$ is charged under ${\rm U}(1)'$. 
The dark sector is allowed to couple to the standard model (SM) sector only through a kinetic mixing between ${\rm U}(1)'$ and ${\rm U(1)}_Y$.
The dark gauge symmetries are assumed to be spontaneously broken, leading to the dark gauge boson masses $m_{A'}$ and $m_{A''}$, respectively. 
We are interested in the scenario where dark gauge bosons are heavier than $\chi_1$ and $\chi_2$.
In this case, $\chi_2$ is disjunct from the SM particles and annihilates only into the other DM species $\chi_1$, while $\chi_1$ can directly annihilate into the SM particles.

The heavy species $\chi_2$ is kept to thermal equilibrium by the assist of the light species $\chi_1$ \cite{Belanger:2011ww} with a thermally-averaged annihilation cross section $\langle \sigma v \rangle_{22\rightarrow 11}$, while $\chi_1$ pair-annihilates directly to SM particles with the cross section $\langle \sigma v \rangle_{11\rightarrow XX}$ where $X$ stands for SM particles.  
We assume $\chi_1$ couples to electrons predominantly, for simplicity. 
Relic abundances of $\chi_1$ and $\chi_2$ denoted as $\Omega_{1}$ and $\Omega_{2}$ respectively, and the total abundance should agree with the observed one, $\Omega_{\text{DM}} = \Omega_1 + \Omega_2 \simeq 0.27$. 
The fraction of $\chi_1$ is expressed by $r_1 = \Omega_1/ \Omega_{\text{DM}}$. 
The free parameters of the model are
$\{ m_1, \, m_2, \, m_{A'}, \, m_{A''}, \, g', \, g'', \epsilon \}$, where $g'$ and $g''$ denote gauge couplings for ${\rm U}(1)'$ and ${\rm U}(1)''$ respectively, and $\epsilon$ is the kinetic mixing parameter between ${\rm U}(1)'$ and ${\rm U(1)}_Y$.
In this paper, we will mainly focus on cosmological features of the model.
Thus, the parameters can be traded with relevant effective parameters $\{ m_1, m_2, \Omega_{\text{DM}}, r_1, \sigma_{\text{self}1}, \sigma_{\text{self}2} \}$, where $\Omega_{\text{DM}}$ is fixed by $0.27$ to yield the observed value and $\sigma_{\text{self}1}$ ($\sigma_{\text{self} 2}$) denotes a self-scattering cross section of $\chi_1$ ($\chi_2$).
We note that $\sigma_{\text{self}1}$ 
(diagram II in Figure \ref{fig:diagram})
plays an important role in sharing the excessive kinetic energy of boosted $\chi_1$ particles with the rest of their species and hence increase the overall $\chi_1$ temperature $T_1 = T_{\chi_1}$ \cite{Kamada:2021muh}. 
In principle, the self-interaction of non-relativistic $\chi_2$ particles can also influence on perturbation evolution, but to focus on the main dynamics of $\chi_1$ self-heating, we will neglect the contribution of $\sigma_{\text{self} 2}$ in our discussions.
Therefore, the parameters determining the most prominent features in cosmological simulation are
$\{ m_1, \, m_2, \, r_1, \, \sigma_{\text{self}1} \}$.
%


\medskip
\noindent {\bf Temperature evolution.}
Cosmological evolution for the number densities of $\chi_1$, $\chi_2$, and SM particles $X$, $n_{1, 2, X}$ are governed by coupled Boltzmann equations as in Refs.~\cite{Belanger:2011ww, Kamada:2021muh}, and will be discussed with details in a companion paper \cite{KKLP}.
Here, we focus on the temperature evolution of the light component $\chi_1$. 

After the decoupling of DM particles from the SM plasma (with temperature $T$), the $\bar{\chi}_2 \chi_2 \rightarrow \bar{\chi}_1 \chi_1$ annihilation 
(diagram I in Figure \ref{fig:diagram})
can leverage the mass gap $\delta m = m_2 - m_1$ to produce energetic $\chi_1$ particles. 
If $\Gamma_{\rm self} > H$,
the excessive kinetic energy ($\Gamma_{\rm heat} > \Gamma_{\rm el}$) is thermalized to increase the overall temperature of $\chi_1$ particles (heating up period in Figure \ref{fig:diagram}). 
The evolution of its temperature $T_1$ is governed by 
\begin{eqnarray}  
\dot{T}_1  \simeq - 2 H T_1 + \gamma_{\text{heat}} T - 2 \gamma_{\chi_1 X} (T_1 - T) \;,
\label{eq:TempEq}
\end{eqnarray} 
where the first term on the right-hand side represents the Hubble friction due to the expansion of the Universe, and the second is responsible for the self-heating with 
$\gamma_{\text{heat}} \simeq {2 n^2_2 \langle \sigma v \rangle_{22\rightarrow 11} \delta m } / \big ( {3 n_1 T} )$ \cite{Kamada:2021muh}.
Note that although the larger mass difference $\delta m$ leads to the higher temperature $T_1$ in general, it is closely associated with the relative number densities of $\chi_1$ and $\chi_2$ as well.
The last term of Eq. (\ref{eq:TempEq}) describes the energy exchange between $\chi_1$ and the SM plasma with
$\gamma_{\chi_1 X} \simeq \big({\delta E}/{ T}\big) n_X \langle \sigma v \rangle_{\chi_1 X}$, 
where $\delta E$ is the change in $\chi_1$ kinetic energy per elastic scattering and $\langle \sigma v \rangle_{\chi_1 X}$ is the thermally averaged scattering cross section of $\chi_1$ off SM particles \cite{Kamada:2021muh}.
\begin{figure}[t!]
    \centering
 \includegraphics[width=1.0\linewidth,clip]{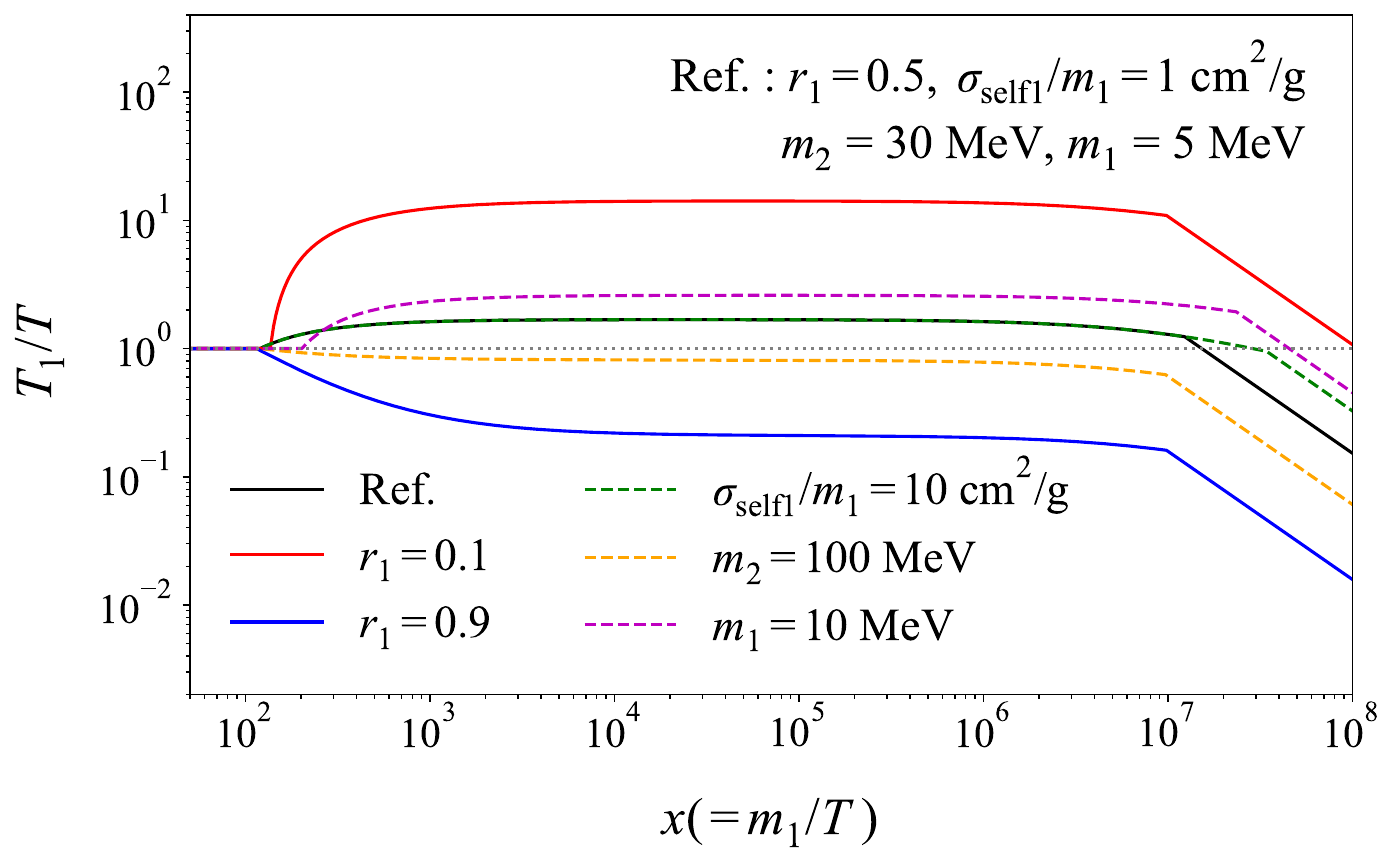}
 \vspace*{-0.3cm}
    \caption{Temperature evolution $T_1$ of dark matter $\chi_{1}$ as a function of $x=m_{1}/T$ under various benchmark model parameters.
    \label{fig:Temperature}}
\end{figure}

Figure \ref{fig:Temperature} shows the evolution of $T_{1}/T$ as a function of $x = m_1/T$ for a few selections of parameter sets by varying one parameter at a time with respect to a benchmark point 
($r_1=0.5$, $\sigma_{\text{self}1}/m_1 = 1~\text{cm}^2/\text{g}$, $m_2 = 30$ MeV, $m_1 = 5$ MeV). 
Results in Figure \ref{fig:Temperature} are exactly what was expected from the illustration in Figure \ref{fig:diagram}.
Several important comments are in order.
First, to boost the temperature $T_1$, a larger $\chi_2$ density is favored ($r_1 \ll 1$).
On the other hand, if the $\chi_2$ density is smaller than $\chi_1$ ($r_1 \to 1$), less amount of energetic $\chi_1$ particles are produced from the $\chi_2$ annihilation. 
With only a small amount of extra energy available, it is not enough to heat a large portion of the relic $\chi_1$ population, which leads to a lower $T_1$ temperature.
Second, the self-interaction $\sigma_{\text{self}1}/m_1$ determines the duration of the self-heating effect. 
For example, with an increase in $\sigma_{\text{self}1}/m_1$ from $1~ \text{cm}^2/\text{g}$ (black solid) to $10~ \text{cm}^2/\text{g}$ (green dashed), the more prolonged time that $T_1$ is heated, and hence its temperature starts to fall much later. 
Finally, the self-heating is more sensitive to the combination of number density $n^2_2/n_1$ than the mass difference $\delta m$. 
An increase of $m_2$ from 30 MeV (black solid) to 100 MeV (orange dashed) leads to a large $\delta m$, but the number density $n_2$ drops as $n_2 \simeq \rho_2 / m_2$ for non-relativistic particles. As a result, $T_1$ of the $m_2 = 100$ MeV case is lower than the $m_2 = 30$ MeV case. 
Similarly, an increase of $m_1$ from 5 MeV (black solid) to 10 MeV (purple dashed) leads to a larger temperature due to the decreased number density $n_1$.


\medskip
\noindent {\bf Perturbation evolution.}
We are now ready to derive the matter perturbation equations.
We focus on the perturbations of $\chi_1$ and $\chi_2$ fluids, while investigating the perturbations of SM fluids is left for future works  \cite{KKLP}. 
We introduce the metric fluctuations in the Newtonian gauge as
\begin{equation}
    ds^2 = -(1 + 2 \Psi)dt^2 + (1 - 2 \Phi) a(t)^2 \delta_{ij} dx^i dx^j \;,
\label{eq:metric}
\end{equation}
where $a(t)$ is a cosmological scale factor, $\Psi$ and $\Phi$ denote scalar perturbations.
Manipulating the standard procedure with the metric fluctuations and covariant derivatives of energy-momentum tensors \cite{baumann_2022}, we derive the following coupled differential equations for the density perturbation (for the first time):
\begin{widetext}
\begin{eqnarray}
    \frac{d \delta_{2}}{dt} + \frac{\theta_{2}}{a} - 3 \frac{d \Phi}{d t} &=& \frac{\big<\sigma v \big>_{2 2\rightarrow 1 1}}{m_{2} \bar{\rho}_{2}} \Bigg( -\Psi \Big( \bar{\rho}^2_{2} - \frac{\bar{\rho}^2_{2,\text{eq}}}{\bar{\rho}^2_{1,\text{eq}}} \bar{\rho}^2_{1} \Big) - \bar{\rho}^2_{2} \delta_{2} + \frac{\bar{\rho}^2_{2,\text{eq}}}{\bar{\rho}^2_{1,\text{eq}}} \bar{\rho}^2_{1} \Big( 2 \delta_{2, \text{eq}} - \delta_{2} -2 \delta_{1 , \text{eq}} + 2 \delta_{1} \Big) \Bigg) \;,  
\label{eq:Perturbation1}\\
\frac{d \theta_{2}}{dt} + H \theta_{2} + \frac{\nabla^2 \Psi}{a} &=& \frac{\big<\sigma v \big>_{2 2\rightarrow 1 1}}{m_{2} \bar{\rho}_{2}}  \frac{\bar{\rho}^2_{2,\text{eq}}}{\bar{\rho}^2_{1,\text{eq}}} \bar{\rho}^2_{1} 
\Big(\theta_{1} - \theta_{2} \Big) \;, 
\label{eq:Perturbation2}\\
\nonumber
\frac{d \delta_{1}}{dt} + \frac{\theta_{1}}{a} - 3 \frac{d \Phi}{d t} &=& -\frac{\big<\sigma v \big>_{2 2\rightarrow 1 1}}{m_{2} \bar{\rho}_{1}} \Bigg( -\Psi \Big( \bar{\rho}^2_{2} - \frac{\bar{\rho}^2_{2,\text{eq}}}{\bar{\rho}^2_{1,\text{eq}}} \bar{\rho}^2_{1} \Big) - \bar{\rho}^2_{2} ( 2\delta_{2} - \delta_{1} ) + \frac{\bar{\rho}^2_{2,\text{eq}}}{\bar{\rho}^2_{1,\text{eq}}} \bar{\rho}^2_{1} \Big( 2 \delta_{2, \text{eq}} + \delta_{1} -2 \delta_{1 , \text{eq}}  \Big) \Bigg) \\ 
&+&  \frac{\big<\sigma v \big>_{1 1\rightarrow X X}}{m_{1} \bar{\rho}_{1}} \Bigg( -\Psi \Big( \bar{\rho}^2_{1} -  \bar{\rho}^2_{1,\text{eq}} \Big) - \bar{\rho}^2_{1} \delta_{1} +  \bar{\rho}_{1,\text{eq}} \Big( 2 \delta_{1,\text{eq}} - \delta_1 \Big) \Bigg) \;,
\label{eq:Perturbation3} \\
\frac{d \theta_{1}}{dt} + H \theta_{1} + \frac{\nabla^2 \Psi}{a} &+& c^2_{s,1} \frac{\nabla^2 \delta_1}{a} =  \frac{\big<\sigma v \big>_{2 2\rightarrow 1 1}}{m_{2} \bar{\rho}_{1}}  \bar{\rho}^2_{2} \Big(\theta_{2} - \theta_{1} \Big)  \;,
\label{eq:Perturbation4}
\end{eqnarray}
\end{widetext}
where $\rho_\alpha$, $\bar{\rho}_\alpha$, $\delta_\alpha = \delta \rho_\alpha / \bar{\rho}_\alpha$, and $\theta_\alpha$ represent energy densities, background energy densities, density contrasts, and velocity divergence fields, respectively.

Eqs. (\ref{eq:Perturbation1}--\ref{eq:Perturbation4}) reproduce the perturbed Boltzmann equations for the CDMs in the vanishing non-gravitational interaction limit. 
The source terms on the right-hand side of Euler equations in Eqs. (\ref{eq:Perturbation2}, \ref{eq:Perturbation4}) describe the momentum transfers between the $\chi_2$ and $\chi_1$ particles which are proportional to the scattering rate, $\big<\sigma v \big>_{2 2\rightarrow 1 1}$ and to the difference in $\chi_2$ and $\chi_1$ velocities, $\theta_2 - \theta_1$.
As expected, the scattering tries to make the $\chi_1$ particles move faster for large $\big<\sigma v \big>_{2 2\rightarrow 1 1}$.
The last term on the left-hand side of Eq. (\ref{eq:Perturbation4}) describes the pressure of $\chi_1$ DM as a consequence of the temperature increase from the self-heating effect.
The sound speed of $\chi_1$ fluid is computed from
\begin{eqnarray}
    c^2_{s,1} = \frac{T_1}{m_1} \Big( 1 - \frac{1}{3} \frac{\partial \ln T_1}{\partial \ln a} \Big) \;,
\label{eq:cs}
\end{eqnarray}
where we have used Eq. (\ref{eq:TempEq}) for $T_1$.
On the other hand, since the $\chi_2$ particles behave like pressureless CDM,
we neglect the pressure term for $\chi_2$ in our analysis.
Finally, the evolution of gravity perturbations, $\Psi$ and $\Phi$, are governed by the
space-time and space-space components of perturbed Einstein equations.
In the presence of the anisotropic tensors due to the contributions from photons and neutrinos, in general $\Psi \neq \Phi$, which we consider properly in our study.
We follow the calculation using two density contrasts in the multi-interacting DM scenario \cite{Becker_2021} with \CLASS~\cite{Blas:2011rf}. 
We use the following parameters based on Planck 2018~\cite{Planck:2018vyg} $\Lambda$CDM parameters, \{$\Omega_{\Lambda}$,\,$\Omega_{m}$,\,$\Omega_{b}$,\,$h$,\,$\sigma_{8}$\} = \{0.6889, 0.3111, 0.049, 0.6766, 0.8102\} for the analysis in the rest of this paper. 

\begin{figure}[t!]
    \centering
     \includegraphics[width=1\linewidth,clip]{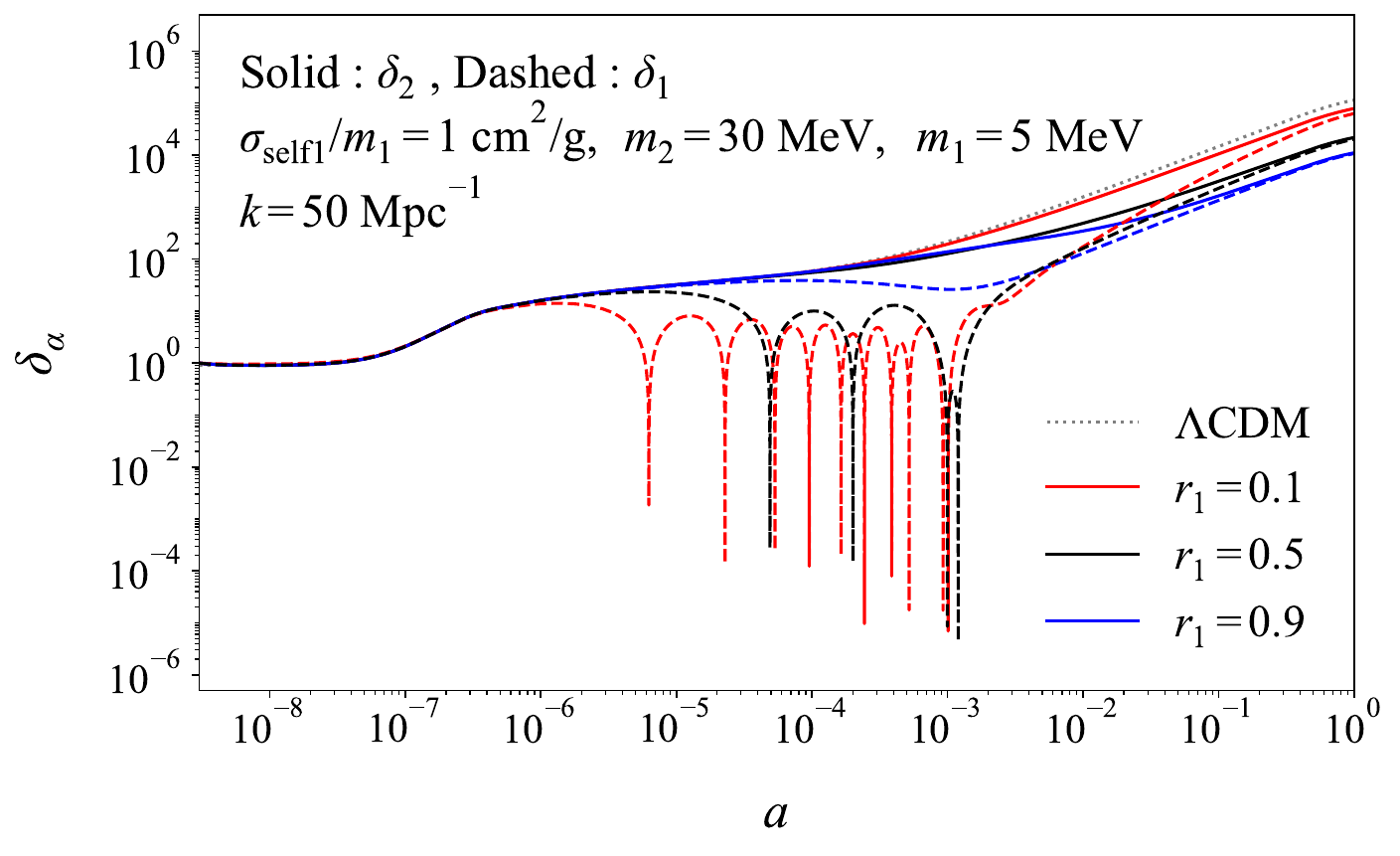}
    \caption{Evolution of dark matter density perturbations under various benchmark model parameters.}
    \label{fig:deltaChi}
\end{figure}

Figure \ref{fig:deltaChi} shows the evolution of density contrasts $\delta_1$ (dashed) and $\delta_2$ (solid) as a function of a scale factor $a$ for a fixed mode $k = 50 ~{\rm Mpc}^{-1}$. 
The $\delta_2$ goes through a significant change around the time of matter-radiation equality, $a \sim {\mathcal O}(10^{-4})$. 
In the single-component DM limit of $r_1 \simeq 0$, the matter energy density is dominated by $\chi_2$, and the evolution of $\delta_2$ resembles that of the single CDM model.
This picture starts to deviate as the $r_1$ increases, where the annihilation cross section $\big<\sigma v \big>_{2 2\rightarrow 1 1}$ becomes large. 
It turns out that in this case the $\delta_2$ experiences a stronger friction due to the disappearance of the gravitational potential well from the $\chi_2$ annihilation, and hence its overall growth is suppressed. 

The $\delta_1$ exhibits the oscillatory behavior which hinders the growth of matter perturbations, even in the deep radiation era. 
In the limit $r_1 \simeq 0$, the annihilation cross section $\big<\sigma v \big>_{2 2\rightarrow 1 1}$ becomes small, which leads to a shallow gravitational potential well of $\chi_1$. 
In this case, the pressure from the self-heating effect dominates over the gravity, causing the oscillatory behavior.
However, as $r_1$ increases, the gravity becomes a match-fit with the pressure, and at some point $\delta_1$ stops oscillating.

All the features described above are somewhat less significant for a small $k$ mode. 
The evolution of $\delta_2$ is nearly the same as that of CDM and unaffected by the ratio parameter $r_1$. 
There is a slight suppression in $\delta_1$ in the limit $r_1 \ll 1$, but it does not show the oscillatory behavior.

\begin{figure}[t!]
    \centering
     \includegraphics[width=0.85\linewidth,clip]{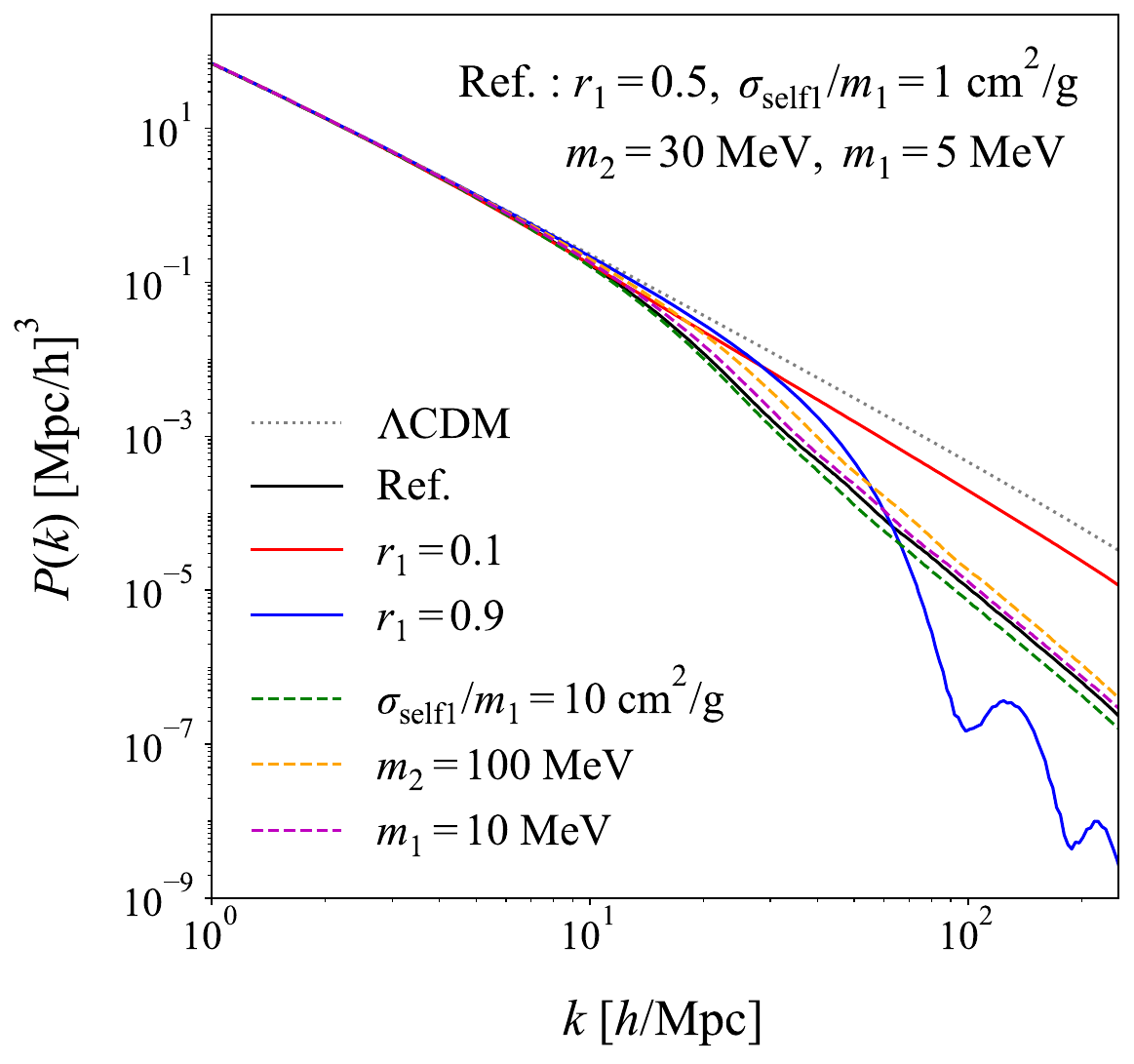}
    \caption{Linear matter power spectrum for various choices of parameters.}
    \label{fig:Linear-PK}
\end{figure}

\medskip
\noindent {\bf Linear power spectrum.}
With the density perturbations $\delta_1$ and $\delta_2$ obtained above, the linear matter power spectrum can be readily computed with \CLASS~\cite{Blas:2011rf}, as shown in Figure \ref{fig:Linear-PK}. 
In the WDM scenarios, the particles are in a relativistic state during the deep radiation epoch, which causes the density perturbations of the particles to be suppressed at the scale below free-streaming length. 
Analogously, in the multi-component model, the power spectrum is suppressed as a result of the self-heating effect through self-interaction of $\chi_1$. 
Therefore, the CDM candidate $\chi_1$ at the MeV mass scale behaves like WDM (at the keV mass scale) via the interactions between dark matter particles. 
Similarly, the oscillatory behavior at $k \gtrsim  100~h/{\rm Mpc}$ (often known as the characteristic of the WDMs) is due to the competition between the gravitational potential and the pressure of the $\chi_1$ fluid. 

The suppression scale $k$ depends on $r_1$. 
For $k < 20 ~h/\rm{Mpc}$, the suppression becomes maximal at an intermediate value of $r_1$, and then the power spectrum increases toward the $\Lambda$CDM case, as $r_1$ increases: e.g., $r_1\sim 0.5$ leads to the lowest power spectrum below $k \sim 20~ h/\rm{Mpc}$ for the reference choice in Figure \ref{fig:Linear-PK}.
On the other hand, for $k > 100 ~h/\rm{Mpc}$, 
the larger $r_1$ leads to the lower power spectrum because there are too many $\chi_1$ particles with self-interaction in the large $r_1$ limit.

Note that the elastic scattering between $\chi_1$ and $\chi_2$, and that between $\chi_1$ and photons contribute to the right-hand side of Eq. (\ref{eq:Perturbation2}) and Eq. (\ref{eq:Perturbation4}), respectively.
The $\chi_1$-photon elastic scattering suppresses the power spectrum as the $\chi_1$ perturbations are suppressed on scales that enter the horizon before the kinetic decoupling \cite{Choi:2015yma,Erickcek:2015jza}.
Such effects appear at $k \gtrsim 500~h/{\rm Mpc}$ for a large value of $r_1$ in our power spectrum, which we find negligible.


\medskip
\noindent {\bf $N$-body simulation.} 
We perform $N$-body simulations using two-component DM simulation~\cite{Todoroki_2018} built on \texttt{GADGET-3}~\cite{Springel_2005,Springel_2008} to investigate the non-linear effects. 
We use a periodic comoving box with the size $3~h^{-1}\rm{Mpc}$  and the total number of particles $128^3$. 
All simulations are carried to the redshift $z_{f} = 0$ with the initial redshift $z_{i} = 49$. 
We adopt \texttt{2LPTIC}~\cite{Crocce_2006} for the second-order Lagrangian perturbation theory to generate initial conditions 
with the linear matter power spectrum as inputs (obtained from \CLASS).
In $N$-body simulations with  WDM~\cite{paduroiu2015structure,Leo_2017,Bode_2001} or mixed CDM-WDM \cite{Maccio:2012rjx,Parimbelli:2021mtp,Boyarsky_2009,Macci__2012,Anderhalden_2012}, the thermal velocity is additionally considered in the Zel’dovich velocity to take into account the thermal effect of WDM particles. 
However, in our simulation, the masses of two particles are large enough compared to WDM (in our model, $m_{1, 2} \geq 5~ \rm{MeV}$), and therefore such thermal effects are negligible. 

\begin{figure}[t!]
    \centering
    \includegraphics[width=0.85\linewidth,clip]{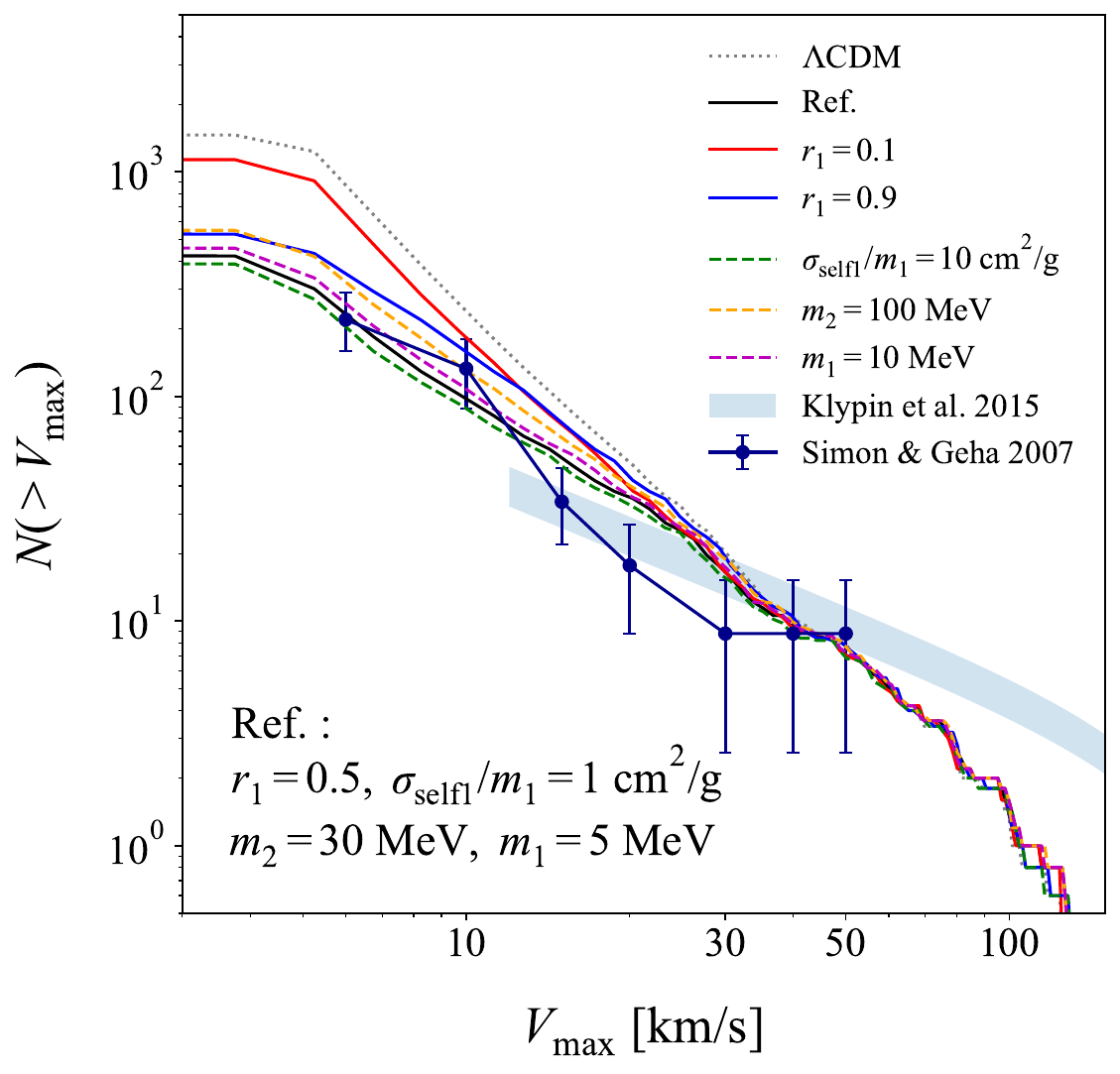}
    \caption{The number of accumulated sub-halos as a function of the maximum circular velocity for various parameters in the two-component dark matter model. 
    \label{fig:MCVF} }
\end{figure}

We extract halo and sub-halo catalogs using Friends-Of-Friends (FOF) halo finder and SUBFIND \cite{Springel:2000qu} algorithms in \texttt{GADGET-4} 
\cite{Springel_2021}. Among many interesting quantities, we focus on the maximum velocity of sub-halos, i.e., MCVF, which is related to the mass scale of the sub-halos. 
Figure \ref{fig:MCVF} shows the number of accumulated sub-halos ($N(>V_{\rm max})$) as a function of their maximum velocity ($V_{\rm max}$) for various selections of parameters with respect to the reference. 
The larger value of $V_{\rm max}$ implies the heavier sub-halos. 
Our results show that in the multi-component scenario, the number of sub-halos is reduced compared to $\Lambda$CDM simulation, mitigating the so-called missing satellite problem \cite{Simon_2007,Klypin_2015}.
Baryonic effects are also known to lower the magnitude of MCVF  \cite{Lovell:2016nkp,Kim:2021zzw}. 
However, such effects may not be strong enough for experimental data, which motivates the inclusion of WDM additionally \cite{Lovell:2016nkp,Kim:2021zzw}. 
Our two-component dark matters are similar in the sense that the light CDM component plays a role of WDM in $N$-body simulation.


\medskip
\noindent {\bf Conclusion and outlook.}
In this paper, we introduced a model with two-component CDM, where the lighter one behaves like WDM. 
For the first time, we have investigated the temperature evolution, density perturbation, linear power spectrum, and maximum circular velocity function via $N$-body simulation. 
Such various cosmological implications provide a crucial input to particle physics experiments. 
For example, performing a simple fit to the MCVF in Figure \ref{fig:MCVF} returns a benchmark reference model, with which one can compute the expected event rates (via the elastic scattering process III in Figure \ref{fig:diagram}) per year at terrestrial experiments such as Super-Kamiokande \cite{Super-Kamiokande:2002weg}, XENONnT \cite{XENON:2017lvq}, and JUNO \cite{JUNO:2021vlw}. 
Further discussion is needed for the full investigation of cosmological and phenomenological implications of boosted dark matter, including nonlinear power spectrum, halo profile, baryonic effect, and gravitational wave signals, which is beyond the scope of the current study and will be performed somewhere else \cite{KKLP}.


\medskip
\noindent {\bf Acknowledgments.}
We thank Donghui Jeong, Jinn-Ouk Gong and Chang Sub Shin for useful discussion. 
The work is supported by the National Research Foundation of Korea (NRF) [NRF-2021R1C1C1005076 (JHK, SHL),  NRF-2019R1C1C1005073 (JCP)].
KK is supported in part by US DOE DE-SC0024407 and University of Kansas General Research Fund allocation. 

\bibliography{bibliography}

\end{document}